\documentclass[doublecol]{epl2}
\usepackage{amsmath}
\usepackage{amssymb}
\usepackage{amsfonts}
\usepackage{graphicx}
\usepackage{dcolumn}
\usepackage{bm}
\usepackage{sidecap}
\usepackage {caption}
\usepackage{float}
\usepackage{parskip}

\title{Impact of heterogeneous delays on cluster synchronization}
\author{Sarika Jalan\footnote{sarika@iiti.ac.in} and Aradhana Singh}
\institute{Complex Systems Lab, Indian Institute of Technology Indore,
M-Block, IET-DAVV Campus Khandwa Road, Indore-452017}
\abstract{
We investigate cluster synchronization in coupled map networks in the
presence of heterogeneous delays. We find that while parity of heterogeneous delays plays a crucial role
in determining the phenomenon of cluster formation, the synchronizability of network
predominantly gets affected by the amount of heterogeneity.
The heterogeneity in delays induces a rich cluster patterns as compared to the homogeneous delays. The complete bipartite networks stands as an extreme example of this richness, where instead of robust ideal driven clusters, versatile cluster 
patterns are observed for which we provide arguments
using the Lyapunov function analysis. 
Furthermore, interplay between the number of connections in the network and the amount of heterogeneity has important role in deciding cluster formation.}

\begin{document}
\maketitle

\subsection{Introduction}
The cluster synchronization has been investigated in many complex systems such as ecological, nervous, 
social, coupled semiconductor lasers and electrical power systems  
\cite{eco_clus,SJ_prl2003,cluster_laser_brain}. 
In these systems the interactions among units are not instantaneous due to the finite speed of 
information transmission causing time delay \cite{book_delay}. 
Most of the work pertaining to delays have considered a homogeneous one, however in real world networks 
rate of information transmission from all the units may not be the same \cite{Neural_mul}. 
Hence model systems
incorporating heterogeneity in delays advances a more realistic framework.
Few previous studies examining systems having heterogeneous delays have shown to
follow emerging behaviors as observed for the homogeneous delays
\cite{hetero_atay}
A recent work demonstrates that an optimal level of delay heterogeneity may maximize the stability of the uniform 
flow which has implications in traffic dynamics \cite{vehicle}. Another recent work involving electronic circuits with heterogeneous delays demonstrates the
change in cluster patterns and suppression of
synchronization \cite{hetero_scholl}.
Furthermore, heterogeneous delays have been shown to bear a more secured communication in chaos based 
encryption systems\cite{optics}.

In this paper we study phase synchronized clusters in the presence of heterogeneous delays.  
We investigate the impact of heterogeneous delays on phenomenon of cluster synchronization.
So far, very few studies have focused on the impact of heterogeneity in delay values on the 
phase synchronized clusters \cite{hetero_scholl, syn_pattern_dis_delay,cluster_laser_brain}. In addition, 
none of the work done so far has attempted to find out the phenomenon behind the cluster synchronization in presence of heterogeneity in delay values.
The undelayed coupled systems have been identified with two different
mechanisms of synchronized clusters formation namely, the driven (D) and the self-organized 
(SO) \cite{SJ_prl2003}. The former refers to the state
when clusters are formed because of inter-cluster couplings, and 
the later refers to the state when clusters are formed because of 
intra-cluster couplings. 

We report that heterogeneity in delay plays a crucial role in the formation
of synchronized clusters as well as the phenomenon behind it. A synchronized cluster, 
in the presence of heterogeneous delays, may be formed because of inter-cluster couplings, 
and may not always be because of a direct coupling between pairs of the nodes in that cluster.
We present results for coupled chaotic maps on various networks namely, 
1-d lattice, small-world (SW), random, scale-free (SF) and the complete bipartite 
\cite{nets-algo, rev_network}.
We observe that large heterogeneity in delay values leads to 
more cluster synchronization and may also lead to a change in the phenomenon behind the cluster formation
 depending on the parity of delay.
For the complete bipartite networks (CBNs) we find that an enhancement in the heterogeneity  
generates a transition from the driven clusters state to a more versatile cluster patterns. 
A cluster pattern refers to a particular synchronized state which contains
information of all the pairs of synchronized nodes distributed in various
clusters in the network \cite{pre_2013}.
\begin{figure}
\begin{center}
\includegraphics[width=\columnwidth]{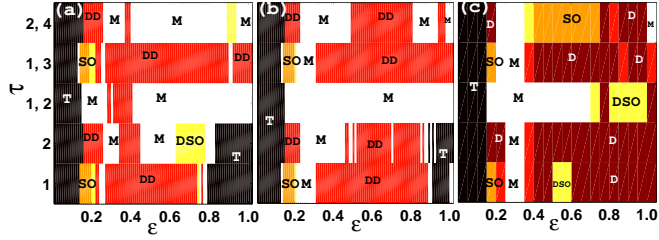}
\end{center}
\caption{(Color online) Phase diagram showing different regions 
in the parameter space of $\tau$ and $\epsilon$ for $f(x)=4x(1-x)$
. Network parameters are $N=500$ and $\langle k \rangle=4$. 
The grey (color) denotes different regions: turbulent (T)(stands for no cluster formation),
ideal driven (D), dominant driven (DD), ideal self-organized (SO), dominant
self-organized (DSO) and mixed (M).
(a), (b) and (c) are for 1-d lattice, SF networks and CBNs respectively.
The figure is obtained by averaging over 20 random initial conditions.
}
\label{Fig-phase}
\end{figure}

\subsection{Model} We consider a network of $N$ nodes and $N_c$ 
 connections between the nodes. Let each node of the network
be assigned a dynamical variable $x_{i}, i=1,2,\hdots,N$. 
The dynamical evolution is defined by the well 
known coupled maps \cite{book_delay},
\begin{equation}
x_i(t+1) = (1-\varepsilon) f(x_i(t)) + \frac{\varepsilon}{\sum_{j=1}^N A_{ij}} \sum_{j=1}^N A_{ij} g(x_j(t - \tau_{ij}))
\label{cml}
\end{equation}
Here $A$ is the adjacency matrix with elements
$A_{ij}$ taking values $1$ and $0$ depending upon whether there is a connection 
between $i$ and $j$ or not. $\varepsilon$ is the overall coupling constant.
The delay $\tau_{ij}$ is the time it takes for the information to reach from a unit $i$ to its neighbor $j$.
The function $f(x)$ defines the local nonlinear map, $g(x)$ defines the nature of coupling between the nodes. 
In the present investigation we consider networks with two types of delay arrangements; (i) 
two delay values and (ii) the Gaussian distributed delay.
The first arrangement is achieved by randomly making a fraction of 
connections $f_{\tau_1}$ conducting with $\tau_1$, and another fraction $f_{\tau_2}$ conducting 
with the delay $\tau_2$. These two parameters are defined as $f_{\tau_1} = N_{\tau_1}/{N_c}$ and 
$f_{\tau_2} = {N_{\tau_2} }/{N_c}$,
where $N_{\tau_1} $ and $N_{\tau_2}$ stands for
the number of connections with delay $\tau_1$ and $\tau_2$, respectively. 
The maximum heterogeneity is exhibited when half of the 
  connections bear $\tau_1$ delay and the other half bear $\tau_2$ delay.
We remark that these definitions do not incorporate the exact values of delay and only takes care of the number of connections conducting with different delay values. 
 
We investigate the first arrangement of two delay values in detail and then consider Gaussian distributed delays arrangement. 
The later arrangement of delay turns out to be a special case of the former.
Note that the delay in the connections are introduced such that $\tau_{ij}=\tau_{ji}$. 
Depending on the parity of delay, we classify three types of the heterogeneity, 
(a) the odd-odd heterogeneity, (b) the odd-even heterogeneity, and (c) the even-even heterogeneity.
We find that these three types have the distinct impact on the coupled dynamics, and hence may
give rise to different patterns of clusters as well as mechanisms behind their origin.

{\it Phase synchronization:}
Nodes $i$ and $j$ are phase synchronized if their minima match all the times in a interval
$T$ \cite{phase_syn}. A cluster of nodes is
phase synchronized if all pairs of nodes of the cluster are phase synchronized.
We use $f_{intra}$ and $f_{inter}$ as measures for intra-cluster and inter-cluster couplings \cite{SJ_prl2003};
$f_{intra}=N_{intra}/N_c$ and  $f_{inter}=N_{inter}/N_c$.
where $N_{intra}$ and $N_{inter}$ are the numbers of intra-cluster 
and inter-cluster couplings, respectively. In 
$N_{inter}$, coupling between two isolated nodes are not included.

\subsection{Coupled maps with two delay values}
Starting with random initial conditions Eq.~(\ref{cml}) is evolved and the phase synchronized clusters 
for $T$ time steps after an initial transient are studied. 
This Letter considers diffusive coupling ($g(x)=f(x)$) because of its relevance in
real world systems \cite{book_delay,hetero_atay}. 
Note that, 
the other forms of the couplings, such as linear, may yield different results for the same
coupling value, but key phenomena observed for diffusive couplings such as
mechanisms of cluster formation would remain same \cite{SJ_prl2003}.
In the following first
we present results for the maximum heterogeneity $f_{\tau_1}=f_{\tau_2}$, followed by the discussions on the
impact of amount of heterogeneity on cluster formation.

\subsection{1-d lattice}
Figs.~\ref{Fig-phase}(a) plots phase diagram depicting different cluster states based on
the values of $f_{inter}$ and $f_{intra}$ for the 1-d lattice. 
For the even-odd parity and $\tau_2=2$ and $\tau_1=1$, weak couplings 
($0.16 \lesssim \varepsilon \gtrsim 0.25$) 
leads to the mixed clusters state. 
As $\varepsilon$ increases, there is an emergence of dominant D clusters 
(Fig.~\ref{Fig-phase}(a)) leading to the mixed clusters for strong couplings.
For odd-odd parity, the ideal SO or the dominant SO clusters 
are formed. 
The ideal SO synchronization refers to a state when
clusters do not have any connection outside the cluster,
except those which are necessary to keep different clusters connected, whereas the ideal D synchronization refers to the
state when clusters do not have any connections within
them and all connections are outside.
The odd homogeneous delay values in this $\varepsilon$ range resembles a similar behavior.
The intermediate and strong coupling
exhibit a manifestation of the dominant D clusters. Comparison with the homogeneous delay evolution leads to the 
conclusion that heterogeneous delays causes an enhancement of synchronization for strong couplings
while keeping the mechanism behind the cluster formation same.

For the even-even parity, the coupled dynamics for the weak $\varepsilon$ range manifests the formation of the ideal D clusters, as observed for the even homogeneous delays (Fig.~\ref{Fig-phase}(a)). With 
increase in the coupling strength fewer number of nodes form clusters, while maintaining 
higher synchrony than the corresponding homogeneous delays. Increase in 
$\varepsilon$ further leads to a transition to the dominant SO
clusters at strong couplings as observed for the homogeneous delays.
The delayed coupled maps on the SW networks generated using Watts-Strogatz algorithm 
\cite{rev_network}, 
do not display any distinguishable changes as compared to the corresponding 1-d lattice described above.

Thus for 1-d lattice and SW networks, change in the parity of heterogeneous delay values 
may give rise to a transition from the D mechanism to SO and vice-versa. We will 
present some understanding of this parity dependence in the section consisting
of the CBNs.

\subsection{SF networks}
We further turn our attention to the SF network, 
which has a completely 
different structural properties \cite{nets-algo} than the 1-d lattice and the SW networks.
The weak coupling range displays a similar result as for the regular networks
described in the previous section for all types of heterogeneity, whereas
intermediate and strong couplings do not display the transition phenomenon
as observed for the regular networks and instead yield the D or mixed mechanism dominant for all
the parities (Fig.~\ref{Fig-phase}(b))
Comparing the three heterogeneity leads to the conclusion that even-even heterogeneity in delays causes 
a less cluster synchronization as compared to the odd-odd and odd-even heterogeneity. The phenomenon
of suppression in synchronization for a particular heterogeneity 
becomes more prominent with the increase in the delay values. 

At strong couplings, heterogeneity in delays manifests more synchronization in SF networks 
as compared to the corresponding 1-d lattice and SW networks 
(Fig.~\ref{Fig-phase}). 
This is not surprising as random networks always display a better synchronization than
corresponding regular networks even for undelayed and homogeneous delays \cite{SJ_prl2003, pre_2013}, the interesting
finding is that this enhancement in synchronization may be accompanied with the nodes
directly connected, which was not observed for homogeneous delays, as evident from
the mixed clusters in Fig.~\ref{Fig-phase}. 

We remark that D clusters were already observed for homogeneous delays in 
intermediate $\varepsilon$ range for 1-d lattice as well as for the SF networks indicating synchronization
between nodes which are not directly connected, therefore occurrence of synchronization
between these nodes for high coupling range does not impart much surprise. We can fairly conclude that phenomenon of SO synchronization has a major role to play in enhancement of
synchronization in the presence of heterogeneous delays, which
further becomes clearly visible for CBN.
In order to understand the copious behavior observed in dynamical evolution of regular and random networks
with heterogeneous delays
we make elaborate investigation for CBNs. The simple structure of 
these networks \cite{nets-algo} on one hand makes analytical studies easier to carry,
on other hand capability of the network to yield rich cluster patterns such as ideal D, SO and mixed clusters
brings it in the same platform of the other random networks. 

\subsection{Complete bipartite networks} 
CBN leads to robust D clusters for the intermediate and strong couplings
for undelayed and homogeneous delays \cite{SJ_prl2003}, where as 
heterogeneity in delays generates the SO, mixed or D clusters depending upon parity and coupling strength in this region 
(Fig.~\ref{Fig-phase}).
 
In the following we perform a Lyapunov function analysis in order to have an understanding of 
destruction of the robust D mechanism for homogeneous delays at the intermediate and strong couplings. 
The Lyapunov function for a pair of nodes on a CBN in the presence of 
heterogeneous delays can be written as:
\begin{eqnarray}
V_{ij}(t+1) &=& [ (1-\epsilon)( f(x_i(t)) - f(x_j(t))) + \nonumber\\
\frac{2\varepsilon}{N}\sum_{j=N/2+1}^N {g}(x_j(t &-& \tau_{ji}))-  
\frac{2 \varepsilon}{N}\sum_{i=1}^{N/2} {g}(x_i(t - \tau_{ij})) ]^{2}
\label{lyap_fun}
\end{eqnarray}
Let us consider a pair of nodes in the same set having the homogeneous delays, which
leads to the situation where coupling terms having delay values in the above equation get canceled,
thereby commencing the D clusters robust against the change in the delay values \cite{pre_2013}. Whereas 
in the presence of heterogeneity in delay values, the coupling term having delay values
does not vanish (Eq.\ref{lyap_fun}), and thus may or may not emulate the synchronization between 
these nodes depending upon the delay arrangements of these two nodes, and
leading to the nodes from the same set organizing into different clusters.
Previous sections show that there is an enhancement in the synchronization
for heterogeneous delays. The CBN stands as an extreme example of 
this enhancement where even the robust D clusters formed for the undelayed and the homogeneous delays 
encounter a transition to the global synchronized state as heterogeneity in the delays are introduced.

Further, using the CBN, we attempt to understand the
parity dependence of the mechanism of cluster formation at weak couplings as observed
for all the network architectures.
A simple analysis for the periodic synchronized state on the CBNs provides
a basic understanding of different behaviors observed for the lower coupling values.
For example, at weak $\varepsilon$ range, the homogeneous delays for $\tau_{1}=1$ manifests the global synchronized state spanning all the nodes
for $0.16 \lesssim \varepsilon \gtrsim 0.2$. 
The dynamical evolution in this range is periodic with periodicity two, say $p1$ and $p2$.
As heterogeneity in the delay values is introduced such that $f_{\tau_1}=f_{\tau_2}=0.5$, 
say at the $(t+1)^{th}$ time step, it leads to the coupling term having delay part in the 
evolution equation for the difference variable of $i^{th}$ and $j^{th}$ nodes as,

\ $f(x_{j}(t-\tau_{2}))-f(x_{i}(t-\tau_{1}))=\left\{ 
\begin{array}{l l}
  0 & \quad \mbox{if $\bigtriangleup \tau=2,4.... $ }\\
  \delta & \quad \mbox{ if $\bigtriangleup \tau=1,3....$}\\ \end{array} \right. $\
  
where $\delta=f(p_1)-f(p_2)$ and $\bigtriangleup \tau=\tau_{2}-\tau_{1}$. $\bigtriangleup \tau$  
is even for the 
odd-odd and the even-even heterogeneity, and odd for the odd-even heterogeneity. Thus the
even-even heterogeneity will retain the behavior followed by the even homogeneous delay values,
and the odd-odd heterogeneity will retain the behavior 
followed by the odd homogeneous delay values. Whereas, the odd-even heterogeneity may disturb 
the behavior manifested by the even homogeneous or odd homogeneous delays.  
Note that for diffusive coupling, the odd delays leads to mismatch in the parity of delay value of the coupling terms, causing  a change in the sign of coupling term. This may cause a significant impact on the dynamics of the coupled system leading to the different phenomena for the odd and even delays \cite{SM}.
 
Next we analyze the origin of SO clusters for the bipartite networks at the
intermediate and strong couplings. A closer look into the time evolution of the coupled nodes in 
the bipartite networks for the intermediate $\varepsilon$ values 
reveals that the heterogeneity suppresses the exact synchronization between the nodes which are
not directly connected while retaining the phase synchronization between them (Fig.1
in \cite{SM}). 
Whereas all the pairs of nodes which are directly connected experience an occurrence of the
phase synchronization producing the global phase synchronized state. 
In order to further explain the synchronization between the nodes from two different sets at 
strong couplings we perform the following analysis.
We consider $\varepsilon=1$, for which all the coupling terms in the difference variable 
for a pair of nodes 
in the same set (i.e. nodes are not directly connected) will get canceled out for the undelayed and 
the homogeneous delayed case, causing to the synchronization of all the pairs of nodes
in the set, whereas the heterogeneous delays do not lead to such a simple 
situation. Let $x_{A}(t)$ being the synchronized dynamics of nodes in the first set and $x_{B}(t)$ 
being the synchronized dynamics of the nodes in the second set. For homogeneous delay,
the difference variable for the 
nodes from the different sets will be:
\begin{equation}
x_{i}(t+1) - x_{j}(t+1) = g(x_{B}(t-\tau)) - g(x_{A}(t -\tau))
\label{diff_var}
\end{equation}
The above equation directs that for the coupled dynamics given by 
a chaotic function (for example, logistic map with $\mu=4$), if the initial conditions for the nodes in two sets are 
different, the difference variable given by \ref{diff_var} will never die. Hence the nodes 
from different sets will never get synchronized ruling out the global synchronized state
for the undelayed and homogeneously delayed case.
However if there exists the heterogeneity in delay values, the difference variable for the nodes 
in the  different sets becomes
\begin{eqnarray}
x_{i}(t+1) - x_{j}(t+1) &=& \frac {2}{N}[\sum_{j=1}^N A_{ij}g( x_{j} (t-\tau_{ij})) \nonumber\\
&-&\sum_{j=1}^N A_{ij}g(x_{i} (t -\tau_{ji}))]
\label{diff_var_hetero}
\end{eqnarray}
For the $g(x)=4 x(1-x)$ at $\varepsilon=1$, the presence of heterogeneity in delay breaks
the restriction \ref{diff_var} and gives rise to a possibility of the synchronization of two 
nodes in the different sets. Though analysis carried out here is done for extreme coupling value ($\varepsilon=1$) and can not be directly applied to other $\varepsilon$ values for which another term
consisting local dynamics of nodes also appears into the difference variable given by
Eq.~\ref{diff_var} and Eq.~\ref{diff_var_hetero}, but having little impact on the dynamical evolution
as compared to the coupled terms for the strong couplings.
\begin{figure}
\centerline{ \includegraphics[width=0.6\columnwidth]{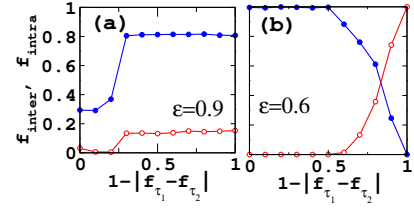}}
\caption{(Color online) Variation of $f_{inter}$ (closed circles) and $f_{intra}$ (open circles) 
as a function of amount of heterogeneity. (a) and (b) CBNs with $N=200$ and $\tau_1=2,\tau_2=4$, (c) and (d)  
SF network with $N=500$ and $\tau_1=1, \tau_2=3$. Both the graphs are for $f(x)=4x(1-x)$}
\label{Fig_D_SO}
\end{figure}

\subsection{Effect of the change in amount of heterogeneity}
So far we have concentrated on the case $f_{\tau_1}=f_{\tau_2}$ corresponding to the 
maximum heterogeneity. We find that
while the amount of heterogeneity plays a crucial role in determining the synchronizability of 
networks, for some cases even demonstrating a transition from no cluster state to all nodes 
forming clusters (Fig.\ref{Fig_D_SO}(a)), the mechanism is still governed by the parity except 
for the CBNs which shows a transition from robust D clusters
state to the SO cluster (Fig.\ref{Fig_D_SO}(b)). We will provide understanding
of this different behavior observed for the CBNs in the later part.

Fig.~\ref{Fig_D_SO}(a) demonstrates clear examples of 
the enhancement in the cluster formation while retaining the mechanism in the presence of the
heterogeneous delays with odd-odd parity. For homogeneous delay (say $\tau=1$), a very less number of 
nodes form clusters (Fig.\ref{Fig_D_SO}). As some connections start conducting with a 
different delay value $\tau_2$, there is no significant 
change in the cluster formation as depicted in the Fig.\ref{Fig_D_SO}.  
 With a further increase in $f_{\tau_2}$, there is an increment in the number of nodes forming clusters reaching to the all nodes forming cluster for 
$1-\mid{f_{\tau_1}+f_{\tau_2}}\mid \gtrsim 0.4$. 

As heterogeneity is demonstrated to always support synchronization, and the CBN 
already displays $100\%$ nodes participating in formation of the robust
D clusters for the homogeneous delay, the only possible way to achieve an enhancement of the synchrony could be via 
synchronization between nodes of two driven clusters 
giving rise to the SO clusters.
The arguments delivered earlier using difference variable 
(Eq.~\ref{diff_var}) directs that more heterogeneity in delays will cause 
more pair of nodes from the same set for which
\begin{equation}
 x_{i}(t+1) - x_{k}(t+1) \neq 0 
\label{diff_var_zero}
\end{equation}
further breaking the restriction on the synchronization between pair of nodes
belonging to different sets, and could be a possible reason behind more 
heterogeneity inducing more SO synchronization.

\subsection{Coupled circle maps}
\begin{figure}
\centerline{\includegraphics[width=0.7\columnwidth]{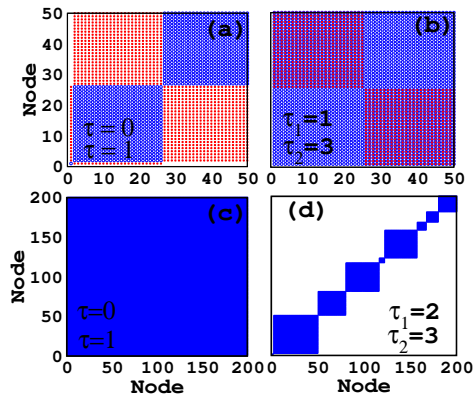} }
\caption{(Color online) Node versus node diagram demonstrating various clusters state
for (a) and (b) for coupled circle maps on CBNs of $N=50$ at $\varepsilon=0.85$, (c) and (d)
for coupled 
logistic maps on globally connected network of $N=200$ at $\varepsilon=1.0$.  
Squares represent clusters and dots imply that the two corresponding nodes are coupled (i.e. $A_{ij}=1$). All the graphs correspond to
$f_{\tau_1}=f_{\tau_2}$.}
\label{Fig_Cluster}
\end{figure}

In order to demonstrate the robustness of the change in the mechanism of cluster formation 
for large amount of heterogeneity in delays at strong coupling range for the CBNs, 
we consider coupled circle maps. Fig.~\ref{Fig_Cluster} plots examples demonstrating the 
transition from the ideal D to the ideal SO cluster state for the
local dynamics defined by circle map \cite{circle_map}, $f(x) = x + \omega + (p/2\pi)sin(2\pi x)$,
with parameters taken in the chaotic regime.  

\subsection{Gaussian distributed delays}
In order to see robustness of the the phenomena, such as enhancement in cluster synchronization and change in the mechanism observed for the CBNs observed for the two delays case, 
we consider the Gaussian distributed delays \cite{SM}, 
and as commented earlier, the Gaussian distributed delays
turns out to be a special case for the former.
 We choose example of SF networks
in order to capture a better overview of the mechanism behind cluster synchronization as 
they are known to exhibit good cluster synchronization for undelayed and delayed evolution.
We find that the distributed delays breaks dominance of any of the two mechanisms,
clearly visible for homogeneous and two delays case, 
leading to the mixed clusters state for $\varepsilon \gtrsim 0.15$.
Comparison with the Fig.~\ref{Fig-phase} indicate that the Gaussian distributed delays
behave in similar manner as the odd-even case of the two delays heterogeneity.

\begin{figure}
\centerline{\includegraphics[width=0.6\columnwidth]{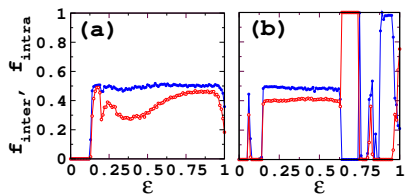}}
\caption{(Color online) Variation of $f_{inter}$ (closed) and $f_{intra}$ (open) circles as a function of $\varepsilon$ for 
SF (left) and CB (right) networks with $N=500$ and for Gaussian distributed delays with mean 
$\bar\tau=5$ and variance $\sigma_{\tau}=2$.}
\label{Fig_Gaussian}
\end{figure}

The other networks, except the CBNs, we have considered, manifest the
similar results as for the SF networks.
The CBNs for the Gaussian distributed delays are capable of displaying all the phenomena of cluster synchronization 
as observed for the two delays case (Fig.~\ref{Fig_Gaussian}(b)), 
leading to the rich cluster patterns depending on the coupling strength.

\subsection{Effect of average degree}
Previous studies demonstrate that undelayed and the homogeneously delayed evolution of all the networks 
with high average degree leads to the global synchronized state after
a critical $\varepsilon$ value, whereas the the introduction of the
odd-even heterogeneity leads to the multi-cluster state.
Note that for this multi-cluster state there is no significant suppression in the overall 
synchronization in the network, as still almost all ($95\%$) the nodes participate in the 
cluster formation. The only difference is that the heterogeneity in delays
breaks the global cluster, distributing its nodes into the different 
clusters (Fig.\ref{Fig_Cluster}).
The Gaussian distributed 
delays at strong couplings also generates the multi-cluster state as observed for the two delays odd-even heterogeneity.
 
For the diffusive coupling we have considered, there is a trade off between the local dynamics and the coupling term.
For strong coupling values, the coupling term dominates over the local dynamics. Again as 
explained earlier, for $\varepsilon=1$, the Lyapunov function for a pair of nodes
(\ref{lyap_fun}) in the globally connected networks would depend only on the term 
$(g(x_j(t - \tau)) - g(x_i(t - \tau)))$ while other terms cancel out. Whereas for the
heterogeneous delays, the Lyapunov function would contain
all the coupling terms, thereby making the stability of the synchronized state dependent on 
the neighbors thereby disturbing the synchronization between the nodes for
the homogeneous delay case.
Therefore, for the heterogeneous delays, a pair of nodes $i$ and $j$ may or may not get synchronized
 depending on the delays connecting to all the neighbors, thereby leading to different cluster
patterns such as multi-cluster state for globally coupled network against global
synchronized state for the homogeneous delays.

\subsection{Discussion and conclusion}
We have studied the impact of heterogeneity in delay values on cluster synchronization
and have presented results for two different delay arrangements; (i) the heterogeneity
with two different delay values, and (ii) the heterogeneity with the Gaussian distributed delays.
For the first case, the cluster synchronization exhibits a dependence on the amount of
heterogeneity in delays. Our results suggest that
the heterogeneous delays accomplishes a better cluster synchronization for 
which we have provided arguments using simple network structures.

Next, we find that at weak couplings the different parities impose 
different constraints on the coupled
dynamics, thereby inducing the different phenomena of cluster formation for which
we have given an explanation 
by considering a simple case of periodic evolution.
For intermediate and strong couplings,
the network properties such as average degree dominates over the nature of delayed
dynamical interaction between the nodes. 
The Gaussian distributed delays exhibit similar
 results as observed for the odd-even
delays displaying the mixed clusters at the weak, intermediate and strong couplings.
All the numerical results indicate that the heterogeneity in delays favors SO mechanism of synchronization
for achieving a better synchrony in network as connections in the network increase. This is more evident in
case of odd-odd heterogeneity, which advances the ideal D clusters for network having less number
of connections and manifests a transition to the SO cluster as connections are
increased.
 Note that for these high average degrees all the
networks (except the CBNs) with homogeneous or
zero delay display the global synchronized
state at strong enough coupling strength, while the networks with
heterogeneous delays yield the multi-clusters state keeping SO mechanism responsible for the
synchronization intact.

Using the Lyapunov function analysis
we have furnished the argument that the heterogeneity in delays
wreaks a different couplings environment for nodes directly connected, which
for strong coupling regime, where coupling term
dominates over the local evolution, being responsible for disrupting the global
cluster. 
We further substantiate that for the CBNs, the heterogeneity in delays
causes a destruction of the robust ideal D clusters and
at very strong coupling where the homogeneous and the undelayed
evolution do not lead to the synchronization between the connected nodes the presence
of heterogeneity in delay brings possibility of synchronization between them.

Further, we find that more heterogeneity in delays is associated with more synchronization.
Thus amount of heterogeneity can be used as a tool to introduce more or less synchronization
in the model networks \cite{cluster_laser_brain}
and can be used to understand versatile cluster patterns observed in real world network \cite{hetero_scholl}. 

To conclude, using extensive numerical simulations accompanied with the
analytical understanding using the Lyapunov function we demonstrate that in the presence of
heterogeneity in delays, the phenomenon for cluster synchronization can be completely
different from the homogeneous delayed evolution. In brain, the time of information
transmission lie in a range exhibiting a heterogeneity in time delay \cite{Neural_mul},
the results presented in this Letter can be used to gain insight into the synchronized activities of
such systems. Furthermore, the heterogeneous delays have been shown to display the regular chaotic patterns in the
brain networks \cite{brain_pattern}. Our results may be further extended to
study the phenomena behind the origin of these patterns.

\subsection{Acknowledgment} We acknowledge 
DST (SR/FTP/PS-067/2011) and CSIR (25(0205)/12/EMR-II) for the financial 
support, and thank complex systems lab members for useful discussions.


\begin{thebibliography}{99}
\bibitem{nets-algo}1-D lattices used in the simulation have circular boundary conditions
with each node having $k$ nearest neighbors. The SW, SF and random networks
are obtained as following \cite{rev_network}: 
SW networks are constructed by rewiring connections of regular 
1-d lattice with a probability $p_r$. SF networks are constructed by starting with 
$\langle k \rangle$ nodes and then adding one node with $\langle k \rangle$ connections 
at each step. 
The random networks are constructed by making connections between all the pairs of
the nodes with probability $p$.
The complete bipartite networks consist of two sets where each node 
of one set is connected with those of the second. Results are presented for
both the sets having equal number of nodes.
\end{thebibliography}
\end{document}